\documentclass[conference]{IEEEtran}
\IEEEoverridecommandlockouts
\usepackage{cite}
\usepackage{amsmath,amssymb,amsfonts}
\usepackage{algorithmic}
\usepackage{textcomp}
\usepackage{xcolor}
\usepackage{comment}
\usepackage{graphicx,pstricks}
\usepackage{graphics}
\graphicspath{{img/}}
\usepackage{booktabs}
\usepackage{tabularx}
\usepackage{tabularx}
\usepackage{xspace}
\usepackage{caption}
\usepackage{array}
\usepackage{cancel}
\usepackage{makecell}
\usepackage{multirow}
\usepackage[caption=false, font=footnotesize]{subfig}
\usepackage[most]{tcolorbox}

\tcbset{
enhanced,
breakable,
left=2mm,
right=2mm,
attach boxed title to top left={
xshift=0.2cm,
yshift=-5mm,
yshifttext=-1mm
},
top=4mm,
colback=orange!5!white,
colframe=orange!100!black,
colbacktitle=orange!100!black,
boxed title style={size=small,colframe=orange!100!black}
}
\usepackage{setspace}
\usepackage{hyperref}

\usepackage{todonotes}

\newcommand{\ie}{\emph{i.e.,}\xspace}
\newcommand{\eg}{\emph{e.g.,}\xspace}
\newcommand{\etal}{\emph{et al.}\xspace}

\definecolor{lightgreen}{rgb}{0.894, 0.961, 0.949}
\definecolor{darkgreen}{rgb}{0.0, 0.576, 0.533}

\tcbset {
  base/.style={
    enhanced,
    breakable,
    arc=0mm, 
    boxrule=0mm,
    colback=lightgreen!20!, 
    top=1mm,
    left=3.5mm,
    leftrule=2mm, 
    right=3.5mm,
    bottom=1mm,
  }
}

\newtcolorbox{mainbox}[1]{
  colframe=darkgreen, 
  base={#1}
}

\def\BibTeX{{\rm B\kern-.05em{\sc i\kern-.025em b}\kern-.08em
    T\kern-.1667em\lower.7ex\hbox{E}\kern-.125emX}}

\begin{document}

\title{\fontsize{23pt}{28pt}\selectfont Human-Written vs. AI-Generated Code: A Large-Scale Study of Defects, Vulnerabilities, and Complexity}


\author{
    \IEEEauthorblockN{Domenico Cotroneo, Cristina Improta, Pietro Liguori}
    \IEEEauthorblockA{University of Naples Federico II, Naples, Italy
    \\\{cotroneo, cristina.improta, pietro.liguori\}@unina.it}
}

\maketitle
\thispagestyle{plain}
\pagestyle{plain}

\begin{abstract}
As AI code assistants become increasingly integrated into software development workflows, understanding how their code compares to human-written programs is critical for ensuring reliability, maintainability, and security. 
In this paper, we present a large-scale comparison of code authored by human developers and three state-of-the-art LLMs, \ie ChatGPT, DeepSeek-Coder, and Qwen-Coder, on multiple dimensions of software quality: code defects, security vulnerabilities, and structural complexity. 
Our evaluation spans over 500k code samples in two widely used languages, Python and Java, classifying defects via Orthogonal Defect Classification and security vulnerabilities using the Common Weakness Enumeration. 
We find that AI-generated code is generally simpler and more repetitive, yet more prone to unused constructs and hardcoded debugging, while human-written code exhibits greater structural complexity and a higher concentration of maintainability issues. 
Notably, AI-generated code also contains more high-risk security vulnerabilities. These findings highlight the distinct defect profiles of AI- and human-authored code and underscore the need for specialized quality assurance practices in AI-assisted programming.
\end{abstract}


\section{Introduction}
\label{sec:introduction}
The emergence of AI-assisted programming has triggered a paradigm shift in software engineering, redefining not only how code is written but also \textit{who}, or rather \textit{what}, is writing it~\cite{ernst2022ai}. 
Large Language Models (LLMs) such as ChatGPT~\cite{chatgpt}, GitHub Copilot~\cite{githubcopilot}, and DeepSeek-Coder~\cite{guo2024deepseek} are no longer limited to generating code snippets or suggesting completions, they now produce entire functions, classes, or even modules with minimal human intervention, transitioning programming from a human-centric design activity into a model-driven process~\cite{lyu2024automatic}. 
According to forecasts by leading AI research labs, by the end of 2025, AI may be responsible for producing up to 90\% of all new code~\cite{anthropics}. 
While this trend brings clear advantages in terms of development speed and accessibility, it also raises critical concerns regarding the quality, correctness, and reliability of automatically generated code.

It is worth noting that this transformation is not merely an evolution of tools and solutions, but a structural redefinition of software development, with the role of developers evolving from designing and implementing code to supervising and adapting model outputs, repositioning them from authors to curators. 
As AI-generated code proliferates, new patterns emerge: code tends to be more concise, heavily pattern-based, and uniform, frequently reusing architectural motifs such as class-based wrappers and utility code. These artifacts are then integrated into production systems, open-source repositories, and developer workflows, often without rigorous inspection. 
As a consequence, AI-generated code is reshaping the stylistic, structural, and quality norms that have traditionally characterized software engineering. 

Unfortunately, current research provides an incomplete picture of the impact of AI code generators on software quality. Most existing studies evaluate AI-generated code in isolation, emphasizing correctness or performance on synthetic benchmarks~\cite{yu2024codereval,mathews2024test}, or considering some aspects of quality~\cite{yeticstiren2023evaluating, liu2024refining} and security~\cite{tihanyi2025secure, fu2023security}.
Few works compare AI-generated code directly with human-authored programs, limiting our understanding of how AI code differs across diverse dimensions such as structural complexity, defect profiles, or vulnerability classes. 
Even fewer studies adopt a standardized framework for evaluating code quality, leading to fragmented and often incomparable findings due to the wide variety of programming languages, tools, and evaluation metrics considered.

These limitations raise a fundamental concern: the frameworks and taxonomies traditionally used to assess software quality are \emph{human-centered}, designed around assumptions about human cognition, error modes, and review processes. 
Yet, AI-generated code introduces new characteristics, including structural redundancy, model-specific vulnerability patterns, or even hallucinations (\eg non-existing libraries or APIs)~\cite{kabir2024stack}. It is unclear whether existing defect classification frameworks can meaningfully characterize these patterns or whether they must be revised and extended to reflect the properties and limitations of AI-generated software.

In this work, we present a large-scale, cross-language comparison of human-written and AI-generated code, focusing on multiple dimensions of software quality: \textit{(i)} code defects, \textit{(ii)} security vulnerabilities, and \textit{(iii)} structural complexity. We focus on Python and Java, two of the most widely used programming languages in both industry and academia~\cite{most-used-languages}, and analyze a large-scale dataset ($>$500k code samples) from human developers ($>$17k GitHub projects) alongside outputs from three popular LLMs: ChatGPT, DeepSeek-Coder, and Qwen-Coder~\cite{hui2024qwen2}. 
We rely on three state-of-the-art static analysis tools to detect issues in the code: Pylint~\cite{pylint} for finding defects in Python functions, PMD~\cite{pmd} for Java, and Semgrep~\cite{semgrep} as a security-oriented solution for discovering vulnerabilities in both. 

To normalize the analysis across tools and languages, we adopt the Orthogonal Defect Classification (ODC) framework by Chillarege \etal \cite{chillarege1992orthogonal}, a well-established method for categorizing defects along independent dimensions such as assignment, algorithm, and interface. ODC allows us to map language-specific defects into a common taxonomy, enabling meaningful cross-language and cross-\textit{author} (\ie human vs. LLMs) comparisons. 

Our study yields the following key findings. First, we reveal systematic differences in defects between AI-generated and human-written code. LLMs tend to produce code that is less structurally complex but more repetitive and prone to specific defect categories, such as variable assignment errors. In contrast, human-authored code exhibits more intricate control flow, algorithmic flaws, and improper exception handling, symptoms often associated with the maintainability and design debt of real-world, mature codebases.
Moreover, compared to software developers, AI models are better at generating Python code yet much worse with Java code ($-$4k and $+$22k defective samples, respectively).
On the security side, we observed that AI-generated code more frequently triggers vulnerabilities associated with high-risk CWE categories, including command injection and hardcoded secrets. 
On average, compared to human code, AI models generate more vulnerable code both in Python ($+$5k samples) and Java ($+$18k samples), fewer lines of code ($-$6.75) and use significantly fewer tokens ($-$63.74), indicating reduced structural and logical complexity and lower lexical diversity.

All code, datasets and results presented in our study are available for transparency and replication purposes on Zenodo~\cite{artifact} and GitHub~\cite{replication}.

In the following, 
Section~\ref{sec:related} discusses related work;
Section~\ref{sec:research_study} describes our research study; 
Section~\ref{sec:dataset} details the dataset construction process; 
Section~\ref{sec:experimental} presents the experimental evaluation;
Section~\ref{sec:threats} discusses threats to validity;
Section~\ref{sec:conclusion} concludes the paper.

\section{Related Work}
\label{sec:related}
Recent studies on AI-generated code have primarily focused on correctness and security~\cite{cotroneo2024vulnerabilities, tihanyi2025secure}, often evaluating just one or two popular models, typically ChatGPT or Copilot~\cite{clark2024quantitative}, on isolated tasks or benchmarks. 
For instance, Pearce \etal~\cite{pearce2025asleep} evaluated Copilot’s security by testing it on 89 coding scenarios and found that around 40\% of its outputs contained vulnerabilities.
Fu \etal~\cite{fu2023security} evaluated code generated by three AI code assistants (\ie GitHub Copilot, CodeWhisperer and Codeium) and performed extensive CWE-based security analysis, including attempts to repair insecure snippets. 
Kharma \etal~\cite{kharma2025security} further highlighted that AI-generated code often fails to adopt modern security best practices, particularly in languages like Java and C++. 
While these studies provide crucial insights into the risks of AI-assisted programming, they are often language-specific, focused solely on security, and lack broader quality metrics.

Beyond security, only a few studies addressed other aspects of code quality. Yeti{\c{s}}tiren \etal~\cite{yeticstiren2023evaluating} benchmarked Copilot, CodeWhisperer, and ChatGPT using HumanEval~\cite{chen2021evaluating} and static analysis tools like SonarQube, revealing wide variability in maintainability and correctness. 
Siddiq \etal~\cite{siddiq2024quality} analyzed real-world developer interactions with ChatGPT, finding that generated code is often significantly modified prior to use due to issues like variable misuse and documentation gaps.
Liu \etal~\cite{liu2024refining} provide an empirical study on ChatGPT-generated Python and Java code, evaluating 4,000 samples for correctness and insights into error types and failure rates stratified by task difficulty.
Improta \etal~\cite{improta2025quality} assessed the quality of a large-scale Python dataset ($>$4M) mined from open-source GitHub repositories using Semgrep and showed that filtering out low-quality functions from training data improves generation. 
Esfahani \etal~\cite{esfahani2024understanding} provide a well-structured defect classification of Python code by using two small AI code generators, \ie CodeT5+ and CodeGen, to generate code for the well-known open-source benchmark, HumanEval. Then, they explore different prompting techniques to automatically fix the problematic code. 

These studies suffer from restricted dataset sizes, single-language or single-model focus, and limited generalizability. 
A key reason for this limitation is that, when dealing with AI-generated code, there is no established taxonomy for systematically classifying software defects. Unlike traditional software engineering, where code evolves within well-defined development lifecycles, AI-generated code is often produced as isolated, context-free snippets, making it difficult to apply process-aware classification frameworks directly.
In conventional systems, one of the most widely adopted models for defect classification is the Orthogonal Defect Classification framework, introduced by Chillarege \etal~\cite{chillarege1992orthogonal}. ODC was explicitly designed to support in-process measurement by categorizing software defects based on their observable impact, independent of language, implementation details, or development stage.

We apply the ODC framework in our study to assess whether it remains a suitable and informative abstraction in this new generation of code development. A recent study by Bogaerts \etal~\cite{bogaerts2024taxonomy} attempted to apply ODC to real-world Python vulnerabilities drawn from CVEs. While their work showed that ODC can bring structure to vulnerability analysis, it also highlighted limitations in using it alone to capture the specificity of modern security issues. 

The most related to our work is the preliminary study by Patel \etal~\cite{patel2024comparative}, which compares Copilot-generated and human-written Java code. However, the study is limited in several ways: it focuses exclusively on a single programming language and two specific software bugs, considers only one AI model, and evaluates a small and narrow dataset of just 90 LeetCode solutions, which lack diversity in scope and style. 

Our study addresses these limitations through a comprehensive and controlled comparison between human-written and LLM-generated code. We analyze human-written code mined from over 17k real-world GitHub projects and code generated by three popular LLMs (\ie ChatGPT, DeepSeek-Coder, and Qwen-Coder) into two widely used languages (\ie Python and Java). 
Our evaluation spans more than 500k code samples and multiple dimensions of code quality, including defect detection, security vulnerabilities, and structural complexity. 
To enable consistent and interpretable comparisons across tools, code authors, and languages, we adopt the ODC framework and MITRE's Common Weakness Enumeration (CWE)~\cite{CWEs}, allowing us to normalize and categorize software issues across languages. 

\section{Research Study}
\label{sec:research_study}

We designed this research study with the aim of answering the following research questions (RQs):
\vspace{0.1cm}

\noindent
$\rhd$ \textbf{RQ$_1$:} \emph{Do defect types and frequencies differ between human-written and AI-generated code across programming languages?}\\
To answer this research question, we systematically compare the distribution of defect types in human-written and AI-generated code across two programming languages. To this aim, we apply two state-of-the-art static analysis tools (\ie Pylint for Python, PMD for Java) and map the resulting defects to standardized Orthogonal Defect Classification categories. This approach allows for a structured comparison of defect types and frequencies among \textit{code authors}, enabling us to quantify differences in code quality between human and AI-generated functions across Python and Java datasets.
\vspace{2pt}

\noindent
$\rhd$ \textbf{RQ$_2$:} \emph{Do security vulnerabilities differ between human-written and AI-generated code across languages, in terms of type and severity?}\\
The goal of this RQ is to investigate and compare security vulnerabilities present in human-written and AI-generated code across Python and Java code. Using Semgrep for static security analysis, we map the identified vulnerabilities to the Common Weakness Enumeration taxonomy and categorize them by severity level. This analysis enables us to assess differences in the types and criticality of security issues introduced by human developers versus AI code assistants, providing insight into potential real-world risks and \textit{author}-specific security behavior.
\vspace{2pt}

\noindent
$\rhd$ \textbf{RQ$_3$:} \textit{Do structural complexity metrics vary between human-written and AI-generated code across programming languages?}\\
The third RQ assesses the structural characteristics of human-written and AI-generated code by analyzing complexity metrics. We evaluate code samples based on quantitative measures such as the number of lines of code, cyclomatic complexity, and token counts, to examine whether AI-generated code shares similar surface-level and deeper structural traits with code written by human programmers.


\subsection{Code Defects Analysis}

\begin{table*}[t]
\centering
\footnotesize
\caption{Summary of the mapping between Orthogonal Defect Classification (ODC) defect types and code smells in Python and Java code. Normal text represents Python rules, while \textbf{bold} text represents Java rules.}
\label{tab:odc_mapping}
\renewcommand\theadfont{\footnotesize\bfseries}
\begin{tabular}{
    >{\raggedright\arraybackslash}m{3.4cm} 
    m{4.7cm} 
    >{\centering\arraybackslash}m{1.1cm} 
    >{\centering\arraybackslash}m{1.1cm} 
    m{5.3cm}
}
\toprule
\thead{ODC Defect Type} & 
\thead{Defect Characteristics} & 
\thead{Python\\Rules} & 
\thead{Java\\Rules} & 
\thead{Example Rules} \\
\toprule
Assignment & Errors in assignment, initialization, or variable binding & 46 & 13 & \texttt{used-before-assignment}, \textbf{\texttt{UnusedAssignment}} \\ \midrule
Algorithm & Logical flaws in computation or data manipulation & 118 & 88 & \texttt{too-many-nested-blocks}, \textbf{\texttt{CyclomaticComplexity}} \\ \midrule
Interface & Issues with interaction between modules, functions, or APIs & 113 & 12 & \texttt{no-value-for-parameter}, \textbf{\texttt{UseProperClassLoader}} \\ \midrule
Checking & Faulty validation or error handling mechanisms & 23 & 26 & \texttt{missing-timeout}, \textbf{\texttt{EmptyCatchBlock}} \\ \midrule
Timing/Serialization & Concurrency, event ordering, or multi-threading defects & 1 & 11 & \texttt{useless-with-lock}, \textbf{\texttt{AvoidSynchronizedStatement}} \\ \midrule
Function/Class/Object & Structural or design errors in function, class, or object organization & 49 & 76 & \texttt{redefined-outer-name}, \textbf{\texttt{MissingOverride}} \\ 
\bottomrule
\end{tabular}
\end{table*}

In this study, we systematically investigate software defects introduced by human developers and AI models when writing code.
A \emph{defect} refers to any violation of correctness, logic, performance or design. These include issues like improper variable initialization, control flow anomalies, or faulty error handling, problems that may lead to program misbehavior but do not necessarily expose the system to external threats.
To detect potential issues, we apply two widely-used static analysis tools: \textit{Pylint}~\cite{pylint} for Python and \textit{PMD}~\cite{pmd} for Java. These tools were selected due to their maturity, extensive adoption in industrial and academic software analysis workflows, and their ability to produce fine-grained, rule-specific outputs. 

Pylint is a deep static analysis tool for Python programs that performs code linting, style checking, bug detection, and design enforcement. It analyzes Python Abstract Syntax Trees (ASTs) to detect a wide range of issues, including syntax errors, type inconsistencies, uninitialized variables, and poor programming practices. Pylint's output is rule-driven, with each rule assigned a unique symbolic name and severity level (\eg convention, warning, error), for a total of 417 detection rules. 
In our setup, we exclude all style-related issues, such as formatting, naming conventions, and documentation warnings, as they do not affect functional correctness, resulting in a final set of 350 rules.

PMD is an extensible static source code analyzer that parses Java code into an AST and applies a wide set of rule-based patterns to detect programming flaws. PMD checks for issues such as unnecessary object creation, performance bottlenecks (\eg avoid instantiating objects inside loops), possible bugs (\eg empty catch blocks, unused variables), and design problems (\eg excessive method complexity). It supports multiple built-in rule sets categorized into best practices, code style, design, documentation, error-prone, multithreading, performance, and security, for a total of 290 rules. 

As before, we excluded from the analysis all rules concerning code style and documentation, as docstrings are stripped from code functions and used as generation prompts (as detailed in \S{}~\ref{sec:dataset}). Additionally, we omitted the security ruleset due to its limited coverage (\ie only two rules), opting instead to incorporate a more comprehensive, security-focused solution later on. In the end, our analysis encompasses 226 detection rules.

To systematically categorize and compare software defects across different code \textit{authors}, \ie human developers or AI code assistants, and across different programming languages, \ie Python and Java, we rely on the \textit{Orthogonal Defect Classification}~\cite{chillarege1992orthogonal}. ODC is a widely recognized method for classifying software defects based on their nature and origin, independent of the specific software development process or application domain. 
It is based on the definition of a set of \textit{attributes}, which are non-overlapping dimensions that measure different aspects of defects, including their type, their impact, their trigger (\ie how the defect was found), and their source. This orthogonality facilitates standardized analysis and defect pattern detection. 

In the paradigm of AI-assisted software development, where developers query LLMs with a short description of the required implementation, the generated code is usually composed of snippets, single functions or short classes, which are then integrated into a larger codebase. For this reason, not all ODC attributes are applicable to this fast-prototyping scenario.

For this reason, we applied the \textit{Defect Type} attribute from ODC to categorize software issues identified in the code. ODC Defect Type describes the nature of the fault independently of specific code details or language constructs, covering eight distinct categories, including Assignment, Algorithm, Interface, Checking, and Timing, among others.

We deliberately excluded other ODC attributes from our analysis, such as the \textit{Trigger} attribute. This dimension captures how a defect is discovered, typically during code inspections, unit testing, integration testing, or system-level testing. However, in our setting, defect detection is performed entirely through static analysis tools applied uniformly across all code samples. This approach does not simulate dynamic software development workflows or staged testing processes in which trigger information would be meaningful. Moreover, LLM-generated code is not compiled, tested, or executed in its original context, making it infeasible to reliably assign trigger labels without speculative or artificial assumptions. By focusing solely on the defect type, we preserve the rigor and comparability of our findings across languages and authors.

For both Python and Java, we manually mapped all the provided detection rules to the most appropriate ODC category. This mapping process was necessary to provide a standardized view of the analysis by normalizing results across different languages and tools, given that static analysis rules are tool-specific and not directly generalizable. Each rule was individually reviewed and assigned an ODC category based on its functional nature and the type of defect it targeted, aiming to minimize ambiguity and maximize consistency across languages.

Given that Pylint and PMD define extensive rule sets with heterogeneous scopes and naming conventions, we performed a detailed manual classification to interpret the intent behind each rule. To ensure the reliability of this process, we adopted a two-stage validation approach. First, two authors independently categorized all rules based on their diagnostic semantics. Disagreements were then resolved through discussion, resulting in a consensus-based mapping. This procedure served as a form of inter-rater agreement, mitigating subjectivity and reinforcing the reproducibility of the classification. This structured resolution of conflicts ensured conceptual alignment with ODC taxonomy. The complete rule-to-category mapping is made available in the replication package for transparency and future extensions~\cite{replication}.

Using ODC allowed us to abstract away from tool-specific rule naming conventions and focus on the broader patterns of defects across datasets. It also enabled cross-comparison between human-written and AI-generated code, highlighting differences in error types rather than being limited to tool output frequencies.

\tableautorefname~\ref{tab:odc_mapping} provides an overview of the categorization. Each ODC Defect Type is described by its defect characteristics, indicating the nature of defects it is intended to capture. For each category, the table reports the number of mapped rules originating from Pylint for Python and PMD for Java. Example rules are provided to show typical violations of each defect type, with Python rules displayed in regular text and Java rules shown in bold.

To make the mapping process more concrete, consider two representative examples. The Pylint rule \textit{used-before-assignment}, which detects the use of variables before initialization in Python, was mapped to the \textit{Assignment} category. This is because such violations reflect fundamental issues in variable binding and initialization, which are key to the definition of assignment-related defects in the ODC framework. On the Java side, the PMD rule \textit{CyclomaticComplexity}, which flags methods with excessive branching logic, was classified under the \textit{Algorithm} category, as it reflects control-flow complexity and potential logical flaws in computation. These examples illustrate how each rule was assigned based on the functional semantics of the issue it detects, rather than tool-specific naming, ensuring consistency in our defect categorization across languages and tools.

Finally, we excluded the \textit{Documentation} and \textit{Build/Package/Merge} defect types from our analysis. 
Documentation-related issues are omitted because docstrings are intentionally removed from the code samples and repurposed as prompts for AI code generation in the dataset construction (\eg \textit{missing-docstring}). 
Similarly, Build/Package/Merge defects pertain to broader integration concerns such as library dependencies, version control, and packaging errors (\eg \textit{import-error}, \textit{NoPackage}), that are not applicable in our context, as we focus exclusively on isolated code snippets rather than complete software systems. As such, these categories fall outside the scope of our study.

Additionally, to ensure compatibility with PMD, which requires valid Java class structures, all Java functions were wrapped in minimal dummy classes prior to analysis. This transformation preserved the original logic while satisfying PMD’s parsing requirements. However, it also introduced artificial class contexts that could inadvertently activate class-level rules unrelated to the function under evaluation. To maintain fairness and analytical consistency, we filtered out all PMD rules targeting class design or object structure (\eg number of fields, method modifiers). This ensured that our evaluation remained focused strictly on function-level logic, the unit of generation for both humans and LLMs, and not on artifacts introduced solely for tool compatibility.
To this end, we excluded 12 additional class-related rules (\eg \textit{UseUtilityClass}, \textit{ClassWithOnlyPrivateConstructorsShouldBeFinal}). Again, all code is provided in the replication package for reproducibility~\cite{replication}.

\subsection{Code Security Analysis}

A \emph{security vulnerability} is a specific type of defect that poses a risk of exploitation, potentially allowing unauthorized access, data leakage, or system compromise. These are classified using the CWE framework and include issues such as command injection, hardcoded secrets, or use of unsafe APIs. 
To detect security vulnerabilities in both Python and Java code, we employ Semgrep OSS~\cite{semgrep}, a modern, lightweight static analysis tool designed for finding security vulnerabilities, correctness bugs, and code quality issues across more than 30 programming languages. 

Unlike traditional static analyzers that often require full code compilation (\eg CodeQL~\cite{CodeQL}, Bandit~\cite{bandit}), Semgrep operates based on syntactic pattern-matching directly on source code, making it particularly well-suited for scanning large-scale, heterogeneous datasets without requiring to build artifacts or full type inference.
Semgrep provides an extensive registry of curated rulesets, each targeting specific families of issues, such as detection of vulnerabilities from MITRE's CWE Top 25~\cite{top25mitre}, unencrypted communication patterns, secrets management flaws, or injection vulnerabilities. 

In this analysis, we focus exclusively on detecting security vulnerabilities using Semgrep rules specifically designed for security analysis. We include all security-focused rules available for Python and Java (as of April 2025). The selected rules cover a broad range of vulnerability categories, including injection attacks, insecure communication, improper authentication, and sensitive data exposure. These rules are sourced from a diverse set of curated Semgrep rulesets targeting well-established security standards such as the OWASP Top Ten~\cite{owasp} and the CWE Top 25. 
To ensure comprehensiveness, we aggregate security rules addressing both general application vulnerabilities and language-specific threats. The complete configuration, including the list of security rules, is available in our replication package~\cite{replication}.

For each identified security issue, Semgrep provides rich metadata, including: \textit{(i)} the rule identifier that triggered the detection; \textit{(ii)} the severity level of the issue (categorized as info, warning, error, or critical); \textit{(iii)} the specific lines of code where the issue occurs; \textit{(iv)} a human-readable explanation of the detected vulnerability; and \textit{(v)} the corresponding mapping to a Common Weakness Enumeration entry. This detailed output enables a systematic and standardized analysis of vulnerabilities across both human-written and AI-generated code samples.

\subsection{Code Complexity Analysis}

To evaluate structural differences between human-written and AI-generated code, we measured several key code complexity metrics that are widely used in software engineering to assess maintainability, readability, and logical depth. Each metric captures a distinct dimension of code structure and computational intricacy, providing a multifaceted view of how code is organized and how difficult it may be to understand, test, or extend. By analyzing these metrics, we aim to identify both quantitative differences between human and AI code, and also stylistic or structural biases in how LLMs approach code generation. 

\noindent
$\blacksquare$ \textbf{Number of Lines of Code (NLOC).} It reflects the overall size of a block of code and can indicate code verbosity or density of implementation. It is computed by counting the total number of lines excluding comments.  

\noindent
$\blacksquare$ \textbf{Cyclomatic Complexity Number (CCN).} It measures the number of independent control paths through a function, thus quantifying logical complexity and the degree of branching or decision-making within the code. It is computed by counting the number of linearly independent paths through the code.

\noindent
$\blacksquare$ \textbf{Token Count (TC).} It captures the number of lexical units (\eg constants, identifiers, operators, keywords) contained in code, providing a finer-grained measure of syntactic richness. 

\noindent
$\blacksquare$ \textbf{Function Name Length (FNL).} It acts as a proxy for naming descriptiveness, with longer names often indicating higher semantic specificity. It is computed as the character count of the extracted function names.

\noindent
$\blacksquare$ \textbf{Unique Tokens (UT).} It is a proxy metric for the vocabulary size. It aggregates the total number of distinct tokens used across the corpus (\ie for a single code author), serving as an indicator of lexical diversity and potentially richer programming expression.

We calculated all complexity metrics using a combination of \textit{lizard}~\cite{lizard} and \textit{tiktoken}~\cite{tiktoken} libraries. We used the former, a lightweight static analysis tool capable of extracting basic structural information from 23 different languages, to extract code snippets. Then, we tokenized the source code using tiktoken, an efficient tokenizer aligned with GPT-4’s encoding schema, to track the average token count and unique tokens per function. We used lizard to automatically calculate NLOC and CCN, while function name lengths were directly measured as the character count of the extracted function names. The metrics were collected separately for each code author (human-written code, ChatGPT, DeepSeek-Coder, and Qwen-Coder) to preserve the granularity of the analysis per source.
Full details of the computation pipeline, along with the scripts used, are available in the replication package~\cite{replication}.

\section{Dataset}
\label{sec:dataset}
\begin{table}[t]
\centering
\caption{Dataset composition.}
\small
\label{tab:dataset_stats}
\begin{tabular}{
>{\raggedleft\arraybackslash}m{1cm} 
>{\raggedleft\arraybackslash}m{1.4cm}
>{\raggedleft\arraybackslash}m{1.4cm}
>{\raggedleft\arraybackslash}m{1.4cm}
>{\raggedleft\arraybackslash}m{1.4cm}
}
\toprule
\textbf{Lang.} & \textbf{\# Samples} & \textbf{\# Repos} & \textbf{Avg. Docstring Len.} & \textbf{Avg. Code Len.} \\
\toprule
\textit{Python} & 285,249 & 12,632 & 25.73 &  
87.21 \\
\textit{Java}   & 221,795 & 4,296 & 24.86 & 79.30 \\
\bottomrule
\end{tabular}
\end{table}

To investigate whether there exist any differences in the defect distribution of human-written and LLM-generated code, we curated a large-scale dataset consisting of both Python and Java code. We selected these programming languages as the target for our study due to their widespread adoption, differing programming paradigms, and relevance in software development~\cite{most-used-languages}. Python, a dynamically-typed, high-level scripting language, is heavily used in AI/ML development. Java, on the other hand, is a statically-typed, object-oriented language widely deployed in enterprise-grade applications, where software quality and security requirements are stringent. By analyzing both, our aim is to ensure that our findings regarding code quality, security vulnerabilities, and defect profiles generalize across different language ecosystems and programming practices. 

We began by adopting and extending the \textit{HMCorp} dataset~\cite{xu2025distinguishing}, which contains 288,508 $\langle$\emph{human-code}, \emph{ChatGPT-code}$\rangle$ pairs in Python and 222,335 in Java.
HMCorp was constructed by filtering the Python and Java subsets of the CodeSearchNet (CSN) dataset~\cite{husain2019codesearchnet} to extract functions authored by human developers. Then, for each function, the corresponding docstring was used to prompt ChatGPT (``gpt-3.5-turbo", April 2023) to generate a matching AI implementation. 
During this filtering process, noisy or malformed samples, such as those containing HTML tags, URLs, or empty function bodies, were systematically removed to ensure dataset quality.
The original CodeSearchNet dataset is a widely-used benchmark comprising $\langle$\emph{documentation}, \emph{code}$\rangle$ pairs in six different languages mined from public, non-forked GitHub repositories sorted by popularity (stars and forks), and is often used to train and evaluate LLMs on code-related tasks~\cite{zheng2023survey, wang2021codet5}. 

However, the released HMCorp dataset excluded the original docstrings and GitHub provenance. To restore this information, we performed a pre-processing step to match each HMCorp sample back to its source in CSN using a combination of function signature and body similarity.
This step allowed us to recover both the documentation and repository metadata, resulting in the creation of complete $\langle$\emph{repository}, \emph{docstring}, \emph{human-code}, \emph{ChatGPT-code}$\rangle$ tuples for each instance. During this process, 3,259 Python and 539 Java samples could not be matched and were discarded, yielding a total of 285,249 Python and 221,796 Java instances.

Then, to further enrich the dataset with more diverse AI-generated implementations, we employed two additional state-of-the-art AI code assistants: DeepSeek-Coder-Instruct (DSC) and Qwen2.5-Coder-Instruct (Qwen). For each sample, we used the original docstring and corresponding function signature as prompts to generate alternative implementations with both models.

\noindent
$\blacksquare$ \textbf{DeepSeek-Coder-Instruct (33B)~\cite{guo2024deepseek}} is part of a series of code language models, each trained from scratch on 2T tokens, with a composition of 87\% code and 13\% natural language. The base model is pre-trained on project-level code corpus by employing a window size of 16K and an extra fill-in-the-blank task. We utilize a quantized, instruction-tuned model with 33 billion parameters, which was further fine-tuned on 2B tokens of instruction data.

\noindent
$\blacksquare$ \textbf{Qwen2.5-Coder-Instruct (32B)~\cite{hui2024qwen2}} is the latest series of Qwen large language models specifically designed for code-related tasks. These models have been pre-trained on an extensive dataset exceeding 5.5 trillion tokens, achieving state-of-the-art performance in code generation, completion, reasoning, and repair tasks. 
We employ a quantized model with 32 billion parameters, which was further trained for instruction-tuning through a multi-stage fine-tuning process.

We followed the default generation settings recommended by each model's authors and used their preferred prompting formats. Example prompts for Python and Java code generation are shown in the following listing.

\begin{mainbox}{}
\label{AI_prompts}
\small
You are an AI programming assistant, utilizing the DeepSeek-Coder model, developed by DeepSeek Company, and you only answer questions related to computer science.

\textbf{\#\#\# Instruction:}
\\

Java

``````\{docstring\}"""

\{signature\} 
\\

\textbf{\#\#\# Response:}
\end{mainbox}

\begin{mainbox}{}
\footnotesize
You are Qwen, created by Alibaba Cloud. You are a helpful assistant.
\\

Python

\{signature\} 

``````\{docstring\}"""
\end{mainbox}

Following generation, a comprehensive code cleanup phase was applied to ensure consistency and validity across all samples. The cleanup process included normalization of indentation, removal of leading or trailing whitespace inconsistencies, and lightweight syntactic parsing checks to detect incomplete or invalid code fragments. In addition, special care was taken to remove any non-code text artifacts often produced by AI models, such as explanations of the code, usage examples, comments unrelated to the function's purpose, or generic model disclaimers. 
To ensure the preservation of valid code, the cleaning procedure was refined iteratively: several rounds of manual inspection were conducted on random subsets of the data to verify that correct code instances were not inadvertently discarded or damaged. Cleaning rules were adjusted as necessary to minimize the loss of properly generated samples. 

Our final dataset consists of 507,044 $\langle$\emph{docstring}, \emph{human-code}, \emph{ChatGPT-code}, \emph{DSC-code}, \emph{Qwen-code}$\rangle$ instances, 285,249 for Python and 221,795 for Java, respectively. The human-written functions span 16,928 unique open-source GitHub repositories, ensuring a representative mix of styles, domains, and development practices.
\tableautorefname~\ref{tab:dataset_stats} provides a full breakdown of the dataset.

\section{Experimental Evaluation}
\label{sec:experimental}
This section illustrates the results of our investigation. First, we dive into the defect categorization and distribution between human-written and AI-generated code across programming languages. We show which are the most occurring ODC defect types and the frequency with which each specific code issue appears in both Python and Java datasets. Then, we focus on the assessment of the security vulnerabilities that are introduced by human developers and AI code generators, respectively. We classify security issues according to MITRE's CWE and analyze their distribution and severity. Finally, we examine whether there are any significant differences in the structure and complexity of Python and Java programs written by developers and generated by LLMs. 

\subsection{Code Defects Assessment}


\begin{table}[t]
\centering
\caption{Defect statistics by language and code author. Blue are \textcolor{blue}{\textbf{best}} scores, red are the \textcolor{red}{\textbf{worst}}.}

\small
\label{tab:defects_summary}
\begin{tabular}{
>{\raggedright\arraybackslash}m{1.5cm}  
>{\raggedright\arraybackslash}m{0.8cm}  
>{\raggedleft\arraybackslash}m{1.5cm}
>{\raggedleft\arraybackslash}m{1.5cm}
>{\raggedleft\arraybackslash}m{1.5cm}
}
\toprule
\textbf{Language} & \textbf{Author} & \textbf{Defective Samples} & \textbf{Incorrect Samples} & \textbf{Total Defects} \\
\midrule
\multirow{4}{*}{\makecell{Python \\(285,249)}}
  & Human       & 158,221 & 3,712  &  429,247 \\
  & ChatGPT     & \textcolor{blue}{\textbf{110,528}} & \textcolor{blue}{\textbf{2,549}}  &  \textcolor{blue}{\textbf{185,110}} \\
  & DeepSeek    & 173,899 & \textcolor{red}{\textbf{20,783}} & 319,643 \\
  & Qwen        & \textcolor{red}{\textbf{178,225}} & 7,190  &  \textcolor{red}{\textbf{574,916}} \\
\midrule
\multirow{4}{*}{\makecell{Java \\(221,795)}}
  & Human       & 76,049 & \textcolor{blue}{\textbf{245}} & 202,129 \\
  & ChatGPT     & 78,241 & 2,056 & 161,056 \\
  & DeepSeek    & \textcolor{red}{\textbf{155,077}} & \textcolor{red}{\textbf{34,382}} & \textcolor{red}{\textbf{242,437}} \\
  & Qwen        & \textcolor{blue}{\textbf{60,389}} & 3,833 & \textcolor{blue}{\textbf{123,180}} \\
\bottomrule
\end{tabular}
\end{table}


To investigate differences in the nature and frequency of code defects across human-written and AI-generated code (\emph{RQ$_1$}), we analyze Python and Java code generated by LLMs (\ie ChatGPT, DeepSeek-Coder, and Qwen-Coder) and compare it to a baseline of human-authored functions. All code samples are analyzed using Pylint (for Python) and PMD (for Java), and the resulting violations are categorized using the Orthogonal Defect Classification framework to ensure consistency across languages and tools. Due to space constraints, in this section we provide an overview of our findings, but complete results can be found in the replication package~\cite{replication}.

\tableautorefname{}~\ref{tab:defects_summary} summarizes the distribution of code issues in terms of \textit{defective samples}, \textit{incorrect samples}, and \textit{total defects}. A sample is considered defective if it contains at least one detected issue; a sample is classified as incorrect if it is either syntactically invalid or entirely missing (\eg empty response due to incorrect interpretation of the input by the model). 
The total number of defects accounts for multiple issues within the same sample, reflecting the cumulative defect count rather than unique instances.

In Python, Qwen-generated code exhibits the highest total number of defects, surpassing both other LLMs and human-written code. This is driven by its high number of defective samples (62.48\%) and a considerable number of incorrect ones (2.52\%).
DSC shows a lower defects rate (60.96\%) but suffers from the highest number of incorrect samples (7.29\%) reflecting more frequent generation failures such as not parseable or incomplete outputs (19,612 empty predictions) and, therefore, potentially less surface for defects.
In contrast, ChatGPT produces the lowest number of defects and defective samples (38.75\%), reflecting a distinct generation style that tends to produce cleaner yet potentially oversimplified code. Also notable, it yields the least number of incorrect samples (0.89\%).

Human-written Python code results in 429,247 total defects, which is higher than both ChatGPT and DSC, but lower than Qwen. This is primarily due to the high number of defective samples (55.47\%) and a relatively low number of incorrect samples (1.30\%). However, as detailed later, a significant portion of these defects ($\sim$47\%) stems from a single code issue detected by Pylint, \ie \textit{protected-access}, which accounts for 202,259 violations. 
This issue refers to accessing class members prefixed with an underscore (\eg \texttt{\_internal\_method()}), which are considered protected by convention. In open-source repositories, especially those involving complex inheritance hierarchies or performance-sensitive designs, such practices are common and often intentional, even if discouraged. If we exclude these violations, the total number of defects in human-written Python code drops to $\sim$227k, making it comparable to ChatGPT. 

In Java, DSC exhibits the worst performance across all dimensions. It produces the highest number of total defects (242,437), the most defective samples (69.92\%), and the most incorrect outputs (15.50\%, 19,071 of which are empty outputs) among all sources, reaching more than 85\% of problematic samples. This suggests that DSC struggles significantly with Java's stricter syntax and structural constraints, even though the most frequent issues refer to best practices and design defects and not critical errors. 
In contrast, human-written Java code results in 202,129 defects across 34.29\% of defective samples, with only 245 incorrect cases, demonstrating higher syntactic and semantic consistency. ChatGPT ranks between these extremes, generating 35.28\% of defective samples. Surprisingly, although being the worst with Python, Qwen shows great capabilities in the generation of Java code, achieving the most robust performance among all authors, with 123,180 total defects, 27.23\% defective samples, and 1.73\% incorrect outputs (3,209 of which are empty).

\begin{figure}[ht]
    \centering
    \subfloat[Python]{%
        \includegraphics[width=1\linewidth]{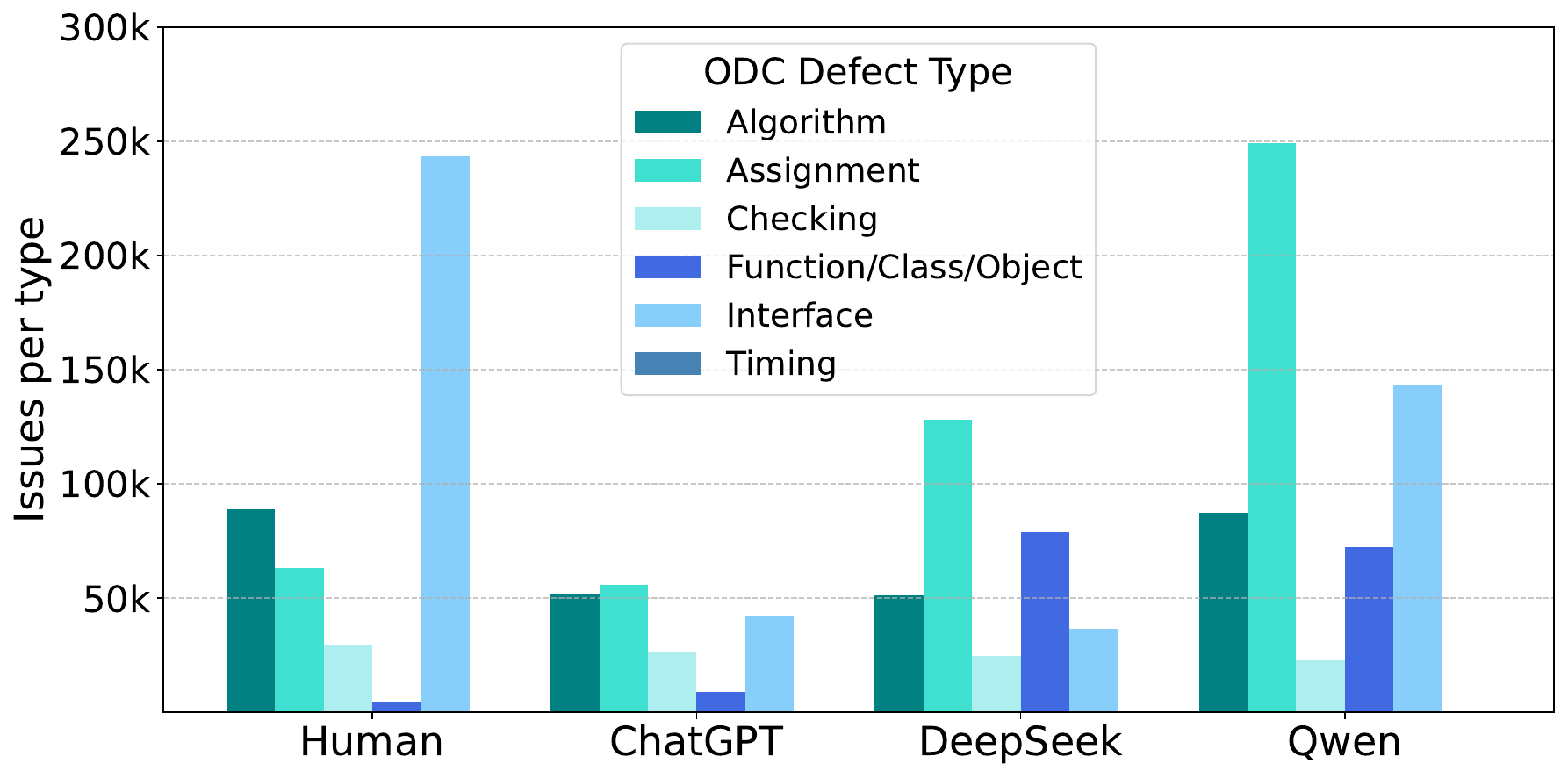}
        \label{fig:python_odc_distribution}
    }\\[1ex]
    \subfloat[Java]{%
        \includegraphics[width=1\linewidth]{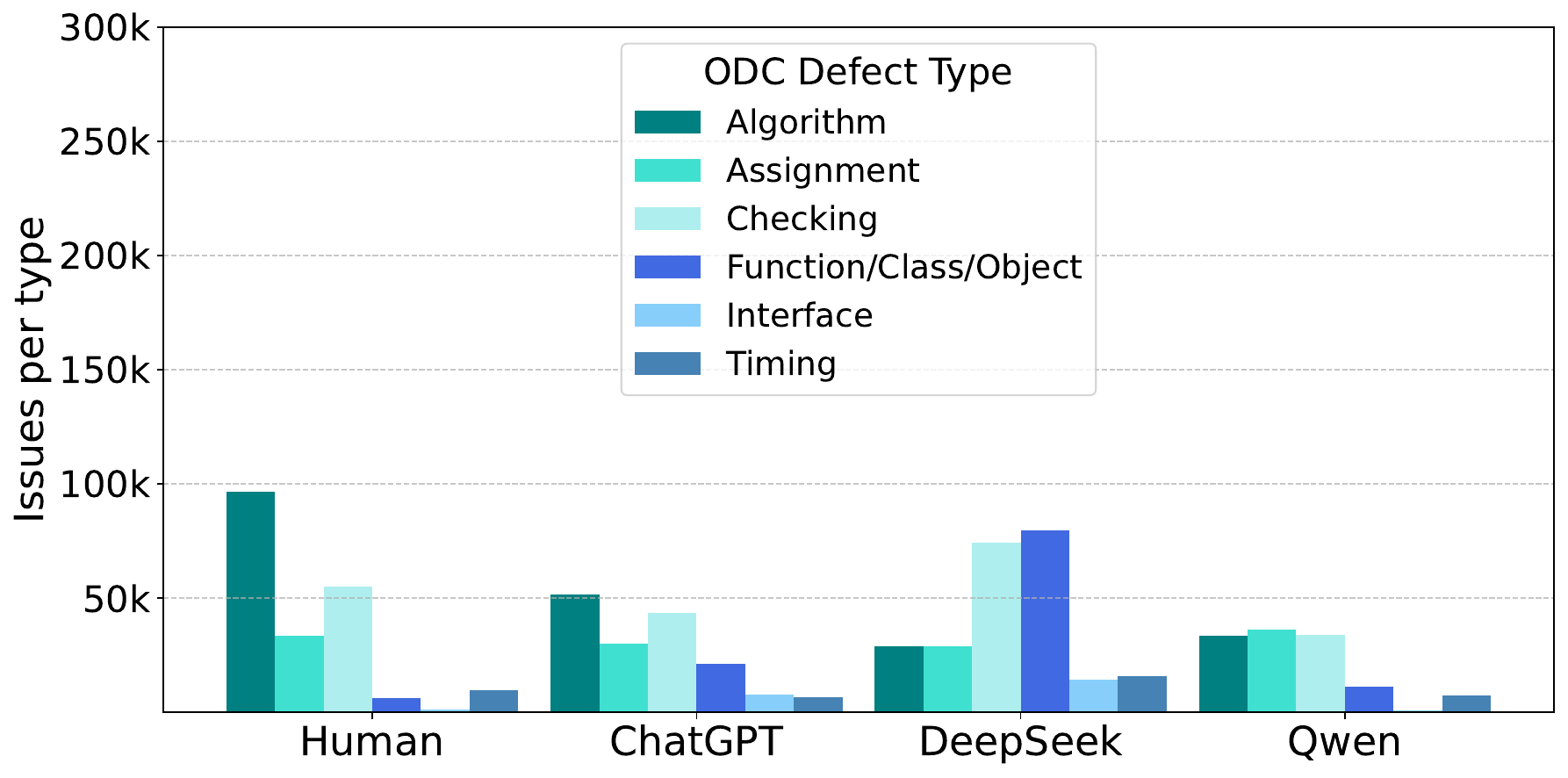}
        \label{fig:java_odc_distribution}
    }
    \caption{Distribution of ODC defect types by code author.}
    \label{fig:odc_defect_distribution}
\end{figure}

To better understand the nature of the issues detected by the static analysis tools, we categorize all reported violations according to the ODC. \figureautorefname{}~\ref{fig:odc_defect_distribution} plots the distribution of ODC defect types across by code author for Python and Java, respectively. Several notable trends emerge.

In Python (\ref{fig:python_odc_distribution}), human-written code is overwhelmingly dominated by interface-related issues, accounting for over 250k violations, more than all other categories combined. This skew is almost entirely due to a protected member being accessed outside a class (\textit{protected-access}). 
The second most frequent type is algorithm due to \textit{no-else-return} (22,619) and \textit{raise-missing-from} (10,108), suggesting unnecessary code and lack of exception traceability, which are typical of hand-written codebases.

In contrast, the top defects in AI-generated code show substantially different patterns. All three models, ChatGPT, DSC, and Qwen, are dominated by assignment issues and the latter two by function/class/object issues, rather than interface and algorithm violations. 
For instance, ChatGPT's most frequent violations are assignment-related (with 32,992 \textit{unused-argument} and 14,408 \textit{unused-variable}), along with \textit{unspecified-encoding} (14,312), indicating failure to declare encoding in file I/O, and \textit{protected-access} (7,108), which is far less frequent than in human-written code. 
DSC displays a unique profile with extremely high counts for \textit{unused-argument} (111,703) and \textit{too-few-public-methods} (64,279), the latter reflecting structural problems in class design, such as the generation of classes with trivial or placeholder content. Qwen, similarly, is dominated by \textit{unused-argument} (213,264), followed by \textit{protected-access} (105,770), and \textit{redefined-outer-name} (57,252), a warning related to overwriting global names within local scopes. 

No timing-related problems are detected apart from 16 \textit{useless-with-lock} issues in ChatGPT code, which fails to establish real locking behavior, potentially leading to thread-safety problems. This is mostly due to Pylint not being well-suited to detect synchronization defects. 

These differences in defects distributions indicate that AI-generated code is often simpler and more repetitive, leading to a higher concentration of variable reuse, shallow argument propagation, and underdeveloped class structures. The recurrence of unused arguments across all AI models in extremely large numbers suggests that they frequently replicate parameter patterns from docstrings or function signatures without correctly using them in the generated body. Meanwhile, structural issues like \textit{too-few-public-methods} and \textit{redefined-outer-name} reflect difficulties in generating coherent and contextually integrated class hierarchies, issue not commonly found in human-written code.


As for Java (\ref{fig:java_odc_distribution}), defect distributions reveal additional differences between human-written and AI-generated code, particularly in how structural and complexity-related issues manifest. Human code exhibits the highest number of algorithmic defects, totaling over 96k, and also shows significant presence in the assignment and checking categories. Indeed, the most occurring defects are \textit{CyclomaticComplexity} (15,435), \textit{AvoidInstantiatingObjectsInLoops} (14,351), \textit{CognitiveComplexity} (13,065), and \textit{AvoidCatchingGenericException} (9,749). 
These reflect code quality concerns that typically arise in real-world, mature codebases, such as overly complex control flow, algorithmic flaws, and improper exception handling. Such defects often relate to maintainability or design debt, rather than outright correctness, and are consistent with the broader, more semantically aware development practices of human programmers.

In contrast, AI-generated Java code, particularly from DSC and Qwen, exhibits very different error profiles. DSC stands out sharply, with high counts for \textit{ImmutableField} (41,669 of function/class/object type) and \textit{SystemPrintln} (37,986), followed by violations such as \textit{UnusedPrivateField}, \textit{AvoidPrintStackTrace}, and \textit{UnusedFormalParameter}. This indicates that DSC-generated code frequently defines fields that could be marked final, overuses low-level debugging constructs like \texttt{System.out.println()} and \texttt{printStackTrace()}, and includes unused or unreferenced fields and parameters, suggesting syntactically plausible but structurally shallow code with limited architectural awareness.

Qwen’s Java outputs exhibit a similar behavior with \textit{UnusedFormalParameter} (24,660) and \textit{SystemPrintln} (13,329) as its most frequent violations. These point again to common problems in low-complexity utility methods and boilerplate code, where unused inputs and basic performance inefficiencies occur frequently.
ChatGPT exhibits a more even spread of defects. Its top violations, \textit{SystemPrintln} (10,428) and \textit{UseVarargs} (9,124), point to best-practice violations rather than complex architectural flaws. Compared to DSC, ChatGPT's Java output is better structured but still presents common coding anti-patterns.

\vspace{2pt}
\noindent
$\blacktriangleright$ \textbf{Key Finding 1:}
\emph{Overall, the qualitative difference between human and AI-generated Java code is clear: human developers tend to write complex, expressive logic that introduces maintainability issues (e.g., cognitive and cyclomatic complexity), while AI code assistants reproduce syntactically valid but semantically shallow code that overuses hardcoded debugging and unused constructs. ChatGPT exhibits a somewhat more stable profile but still deviates from human practices, particularly in areas like method scoping and parameter usage. On average, compared to human code, AI models are better at generating Python code yet much worse with Java code ($-$4k and $+$22k defective samples). These findings further reinforce the need to evaluate AI-generated code not only for correctness but also for design-level quality.}

\subsection{Code Security Assessment}

\begin{table}[t]
\centering
\caption{Security vulnerabilities statistics by language and code author. Blue are \textcolor{blue}{\textbf{best}} scores, red are the \textcolor{red}{\textbf{worst}}.}
\small
\label{tab:vulnerabilities_summary}
\begin{tabular}{
>{\raggedright\arraybackslash}m{1.5cm}  
>{\raggedright\arraybackslash}m{0.8cm}  
>{\raggedleft\arraybackslash}m{1.5cm}
>{\raggedleft\arraybackslash}m{1.5cm}
>{\raggedleft\arraybackslash}m{1.5cm}
}
\toprule
\textbf{Language} & \textbf{Author} & \textbf{Vulnerable Samples} & \textbf{Unique CWEs} & \textbf{Total CWEs} \\
\midrule
\multirow{4}{*}{\makecell{Python \\(285,249)}}
  & Human       & \textcolor{blue}{\textbf{15,835}} & \textcolor{red}{\textbf{45}} & \textcolor{blue}{\textbf{25,678}} \\
  & ChatGPT     & 22,864 & \textcolor{red}{\textbf{45}} & \textcolor{red}{\textbf{40,035}} \\
  & DeepSeek    & \textcolor{red}{\textbf{22,878}} & \textcolor{blue}{\textbf{43}} & 39,982 \\
  & Qwen        & 16,316 & \textcolor{red}{\textbf{45}} & 34,250 \\
\midrule
\multirow{4}{*}{\makecell{Java \\(221,795)}}
  & Human       & \textcolor{blue}{\textbf{ 6,660}} & 42 & \textcolor{blue}{\textbf{11,677}} \\
  & ChatGPT     & 18,144 & 49 & 29,807 \\
  & DeepSeek    & \textcolor{red}{\textbf{43,386}} & \textcolor{red}{\textbf{51}} & \textcolor{red}{\textbf{76,678}} \\
  & Qwen        & 13,620 & \textcolor{blue}{\textbf{41}} & 21,301 \\
\bottomrule
\end{tabular}
\end{table}

\begin{figure}[ht]
    \centering
    \subfloat[Python]{%
        \includegraphics[width=1\linewidth]{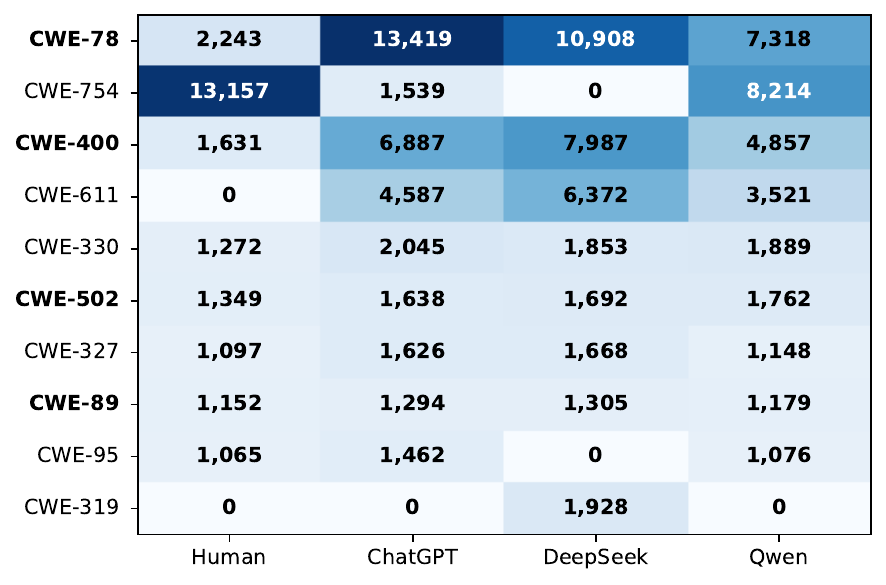}
        \label{fig:heatmap_python}
    }\\[1ex]
    \subfloat[Java]{%
        \includegraphics[width=1\linewidth]{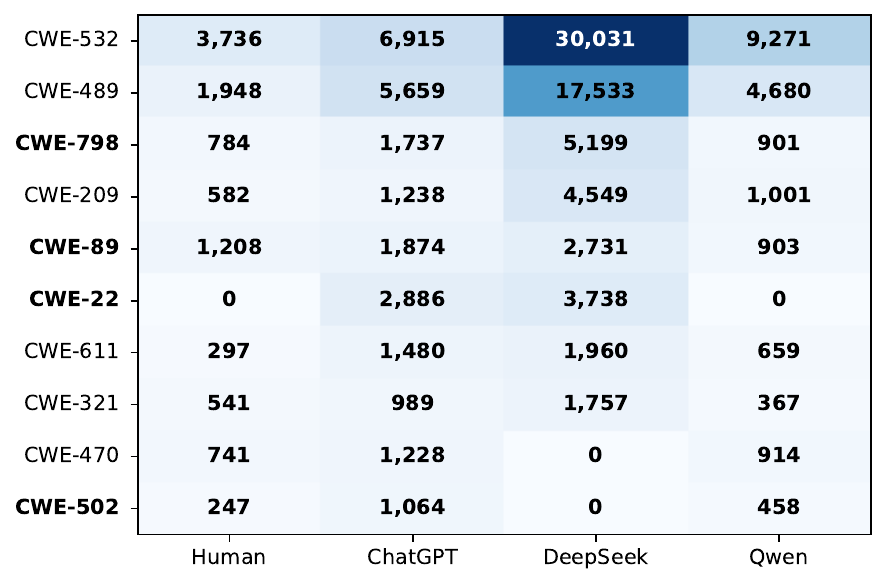}
        \label{fig:heatmap_java}
    }
    \caption{Heatmap of top-10 CWEs distribution across code authors.}
    \label{fig:heatmap_CWEs}
\end{figure}

Although AI code assistants can reliably generate syntactically correct and functionally valid code, recent evidence suggests that they may also propagate insecure coding patterns, often in subtle and systematic ways. Understanding whether the nature of these vulnerabilities diverges from defects traditionally introduced by human developers is essential to adapting software security tools and mitigation frameworks accordingly.

To address \emph{RQ$_2$}, we used Semgrep, configured with an extensive set of security detection rules mapped to the CWE, to extract issue counts, severity levels, CWE types, and vulnerability distributions for each code sample.
This analysis enables a structured comparison of both the frequency and nature of security issues, with the aim of determining whether AI-generated vulnerabilities form a distinct risk profile. The complete results are available in the replication package~\cite{replication}.

\tableautorefname{}~\ref{tab:vulnerabilities_summary} provides a summary of the frequency and \textit{density} (\ie nr. of issues per sample) of security vulnerabilities detected. We report the number of vulnerable samples (\ie unique samples containing at least one issue), of unique CWE identifiers triggered, and the total number of issues detected across all samples (\ie a single function may contain several different vulnerabilities).

In Python, AI-generated code exhibits a higher volume but equal coverage of vulnerabilities than human-written code. DSC produces the most insecure code, with 22,878 vulnerable samples and 39,982 total CWE-triggered issues, followed closely by ChatGPT (22,864 vulnerable samples and 40,035 issues), reaching $\sim$8\% of the dataset. 
Both models substantially exceed the human baseline of 5.55\% vulnerable samples and 25,678 total issues. Qwen shows a somewhat more conservative behavior (5.72\% vulnerable samples and 34,250 issues), though still higher. Interestingly, despite these large disparities in total issue count, the number of distinct CWEs triggered by each AI model in Python is nearly identical to human code (45 for humans, ChatGPT and Qwen, 43 for DSC), suggesting that AI-generated code exhibits not only more vulnerabilities but also repeated instantiations of the same types of flaws.
Moreover, while human-written code contains 1.62 issues on average, this density reaches 1.75 for ChatGPT and DSC, and 2.10 for Qwen. 

In Java, the gap is even more pronounced. Code generated by DSC stands out with an extremely high number of vulnerable samples (19.56\%) and total issues (76,678), triggering 51 distinct CWE types, the highest among all authors and languages. This indicates both a broader vulnerability surface and more frequent violations. ChatGPT and Qwen also exhibit numerous vulnerable samples (8.18\% and 6.14\%, respectively) compared to the human baseline of 3\%, with a similar or higher CWE variety (49 and 41 vs. 42 for humans). Overall, DSC consistently demonstrates the worst security profile in both languages, while human-written code is consistently the most secure across all metrics, followed by Qwen being the least prone to generate insecure code. 

\figureautorefname~\ref{fig:heatmap_CWEs} presents detailed heatmaps illustrating the distribution of the ten most frequent CWEs in Python and Java, respectively, across different code authors. CWEs belonging to MITRE's Top 25 most dangerous software weaknesses~\cite{top25mitre} are shown in bold.
In Python (\ref{fig:heatmap_python}), AI-generated code disproportionately triggers CWE-78 (OS Command Injection), with DSC and ChatGPT showing 10,908 and 13,419 instances respectively, compared to only 2,243 in human code. 
A similar trend is observed for CWE-400 (Uncontrolled Resource Consumption) and CWE-611 (Improper Restriction of XML External Entity Reference), which are almost absent in human-written code but occur in thousands of cases in DSC and ChatGPT outputs.
Notably, these two models exhibit highly similar insecure behavior in Python, both in terms of volume and distribution, indicating that despite model differences, they may internalize and reproduce comparable insecure generation patterns. 
Human-authored Python code, by contrast, is heavily skewed toward CWE-754 (Improper Check for Unusual or Exceptional Conditions), with 13,157 instances, indicating a different defect profile and confirming the frequent lack of exception handling in projects written by human programmers.

DSC further distinguishes itself by generating code vulnerable to CWE-319 (Cleartext Transmission of Sensitive Information), which does not appear in any other author's code. These gaps emphasize that while AI models broadly follow similar trends, certain failure modes are model-specific and need to be addressed differently from traditional human-written insecure code.

Java results (\ref{fig:heatmap_java}) reveal a similar divergence in defect distributions. DSC again produces the highest frequencies across several high-impact CWE types. For instance, its code accounts for 30,031 instances of CWE-532 (Information Exposure Through Log Files), compared to 3,736 for humans. It also dominates in CWE-489 (Active Debug Code), CWE-798 (Use of Hardcoded Credentials), and CWE-209 (Information Exposure Through an Error Message), all of which are significantly underrepresented in human-written code. ChatGPT and Qwen show similar but less extreme trends, while human-authored Java code displays a flatter distribution with fewer high-impact CWEs overall.

The increased number of unique CWEs in AI-generated code reflects a broader attack surface and a tendency to reproduce known insecure patterns more frequently. Notably, many of the most prevalent vulnerabilities align with categories listed in MITRE’s Top 25, underscoring their severity. CWE-78 (OS Command Injection), CWE-400 (Uncontrolled Resource Consumption), and CWE-798 (Use of Hardcoded Credentials) are particularly frequent in code produced by ChatGPT and DSC, but rare in human-written code. 

\vspace{2pt}
\noindent
$\blacktriangleright$ \textbf{Key Finding 2:}
\emph{In summary, these findings confirm that AI-generated code differs from human-written code not only in the number of security vulnerabilities but also in their nature and distribution. AI models are particularly prone to recurring instances of high-severity issues such as injection flaws, insecure configurations, and information exposure. These trends hold across both Python and Java, although they are more extreme in the latter.
On average, compared to human code, AI models generate more vulnerable code both in Python ($+$5k samples) and Java ($+$18k samples). The combination of higher defect volume, greater CWE diversity, and a skew toward high-severity categories underscores the importance of properly assessing the security risks of AI-generated code.}

\subsection{Code Complexity Assessment}

\begin{table}[ht]
\centering
\caption{Comparison of structural complexity metrics. \textbf{Bold} values indicate the highest score per metric.}
\small
\label{tab:complexity_table}
\begin{tabular}{
>{\raggedright\arraybackslash}m{1cm}  
>{\raggedright\arraybackslash}m{0.8cm}  
>{\raggedleft\arraybackslash}m{0.8cm}
>{\raggedleft\arraybackslash}m{0.8cm}
>{\raggedleft\arraybackslash}m{0.8cm}
>{\raggedleft\arraybackslash}m{0.8cm}
>{\raggedleft\arraybackslash}m{1cm}
}
\toprule
\textbf{Language} & \textbf{Author} & \textbf{Avg. NLOC} & \textbf{Avg. CCN} & \textbf{Avg. TC} & \textbf{Avg. FNL} & \textbf{UT} \\ 
\midrule
\multirow{4}{*}{Python} 
& Human         & \textbf{12.72} & \textbf{3.97} & \textbf{142.45} & 14.16 & \textbf{70,542} \\ 
& ChatGPT       & 6.89 & 2.47 & 79.33 & \textbf{17.56} & 62,717 \\ 
& DeepSeek & 5.16 & 2.00 & 68.37 & 12.51 & 60,199 \\ 
& Qwen     & 4.47 & 1.84 & 58.68 & 13.95 & 63,265 \\ 
\midrule
\multirow{4}{*}{Java} 
& Human         & \textbf{13.38} & \textbf{3.48} & \textbf{119.66} & 13.98 & \textbf{57,356} \\ 
& ChatGPT       & 8.42 & 2.35 & 76.51 & 18.92 & 52,540 \\ 
& DeepSeek & 6.70 & 1.83 & 59.60 & \textbf{19.32} & 47,737 \\ 
& Qwen     & 6.14 & 1.90 & 61.43 & 13.85 & 50,698 \\ 
\bottomrule
\end{tabular}
\end{table}

To answer \emph{RQ$_3$}, we analyzed structural complexity metrics, including the number of code lines (\texttt{NLOC}), the cyclomatic complexity (\texttt{CCN}), average and unique token counts (\texttt{TC} and \texttt{UT}), and function name length (\texttt{FNL}), across human-written and AI-generated code samples. 
The goal is to both quantify differences in surface-level complexity and also to understand whether AI-generated code mirrors the structural richness and diversity observed in human-written software. 
\tableautorefname{}~\ref{tab:complexity_table} summarizes the results.

In Python, the structural gap between human-written and AI-generated code is particularly pronounced. Human-written Python functions average 12.72 NLOC and exhibit a CCN of 3.97 on average, reflecting deeper logical branching and control flow compared to AI-generated counterparts. In contrast, code snippets generated by ChatGPT, DSC, and Qwen are substantially shorter and simpler, with average NLOC values of 6.89, 5.16, and 4.47 respectively, and CCN values of 2.47, 2.00, and 1.84. The average length of human Python code (142.45 tokens per function) is markedly higher than in AI-generated functions, with the most concise code produced by Qwen (58.68 tokens). Finally, AI-generated Python functions tend to have comparable or slightly longer function names relative to human code, with ChatGPT being the most prolix.

In Java, a similar but slightly less extreme trend emerged. Human-written Java functions demonstrate significantly higher scores across all metrics: the average NLOC for human functions is 13.38, compared to 8.42 for ChatGPT, 6.70 for DSC, and 6.14 for Qwen. Cyclomatic complexity is also higher in human Java code (3.48) relative to AI-generated output, which ranged from 1.83 to 2.35. Token counts follow the same pattern, with human-written Java averaging 119.66 tokens per function, while AI code ranges between 59.60 and 76.51 tokens. Interestingly, AI-generated Java functions exhibit notably longer average function names, with DSC (19.32 characters) and ChatGPT (18.92) exceeding the average function name length chosen by human developers of 13.98, suggesting a pattern of more verbose or explicitly descriptive naming conventions in AI outputs.

In addition to conventional structural metrics, we also analyzed the unique token count for each author, representing the number of distinct lexical units (\eg keywords, identifiers, operators) used across the code samples. Vocabulary diversity serves as a proxy for syntactic richness and expressiveness: a higher number of unique tokens may indicate a broader use of language features, more varied code patterns, and potentially more nuanced or sophisticated program logic. Conversely, a low unique token count suggests more repetitive, template-like code structures. In both Python and Java, human-written code exhibited significantly higher unique token counts than AI-generated code, with 70k unique tokens in human Python samples compared to 62k, 60k and 63k for ChatGPT, DSC and Qwen, and 57k unique tokens in human Java compared to 52k, 47 and 50k for AI counterparts. These findings suggest that, especially when writing Python code, human developers employ a richer and more varied programming style, while AI models, despite generating syntactically valid code, tend to reuse a narrower subset of language constructs. 

\vspace{2pt}
\noindent
$\blacktriangleright$ \textbf{Key Finding 3:}
\emph{Overall, the results consistently demonstrate that human-written code tends to be structurally more complex, both in terms of size and logical structure, across both programming languages. In contrast, AI-generated code, particularly that produced by DSC and Qwen, favors brevity and reduced complexity. On average, it contains fewer lines of code ($-$6.75), exhibits lower cyclomatic complexity ($-$1.66), and uses significantly fewer tokens ($-$63.74), indicating reduced structural and logical complexity. Additionally, AI-generated code tends to contain fewer unique tokens ($-$7,756), suggesting lower lexical diversity. These findings suggest that AI models prioritize generating compact code over producing highly structured or richly branched functions. While reduced complexity and lexical variety may offer benefits such as improved readability, it may also introduce risks of under-specification, limited error and edge cases handling, or shallow logic coverage.}



\section{Threats to Validity}
\label{sec:threats}

\noindent
\textbf{{Construct Validity}:}
Our study exclusively relies on static analysis tools for detecting code defects, which inherently limits our evaluation to syntactic and structural aspects, excluding runtime behaviors, context-specific errors, and deeper semantic correctness. This limitation arises because the code generated by LLMs often comprises standalone functions or snippets lacking executable context, making dynamic approaches such as runtime verification and testing impractical. Nevertheless, the use of mature and widely adopted static analysis tools like Pylint and PMD ensures consistent, meaningful defect detection, effectively addressing our research goal of comparative quality assessment between human and AI-generated code.
Moreover, given the heterogeneous nature of rule definitions across static analyzers like Pylint and PMD, mapping them to ODC categories inevitably involves interpretive judgment. To mitigate potential subjectivity, we employed a structured two-stage validation protocol where two authors independently classified each rule, followed by a consensus resolution of any disagreements. While this process enhances reliability and reproducibility, some degree of ambiguity remains, particularly for rules that straddle multiple defect types. Future work could explore automated or learning-based classification approaches to further improve consistency.


\noindent
\textbf{Internal Validity:}
Our dataset builds upon the publicly available and peer-reviewed \textit{HMCorp} dataset~\cite{xu2025distinguishing}, which includes over 500k human-written and ChatGPT-generated code samples in Python and Java. While this provides a strong baseline, potential biases could be introduced during our dataset expansion phase. Specifically, we generated additional AI code using two state-of-the-art models, \ie DeepSeek-Coder-Instruct and Qwen2.5-Coder-Instruct, based on the original docstrings and function signatures. Although the generation process was standardized to ensure consistency across models, discrepancies in model behavior and prompt interpretation could introduce subtle variability. Nonetheless, we mitigate this threat by \textit{(i)} incorporating outputs from multiple LLMs to increase diversity, \textit{(ii)} applying the same prompting templates across models to reduce bias, and \textit{(iii)} anchoring our study on a high-quality and publicly available dataset that meets modern standards for scale, diversity, and cross-language support.

\noindent
\textbf{External Validity:} Our findings' external validity may be constrained by the selection of programming languages (Python and Java), the specific LLMs used (ChatGPT, DeepSeek-Coder, Qwen-Coder), and our prompt-based generation strategy relying primarily on docstrings. While these selections represent practical and widely used configurations in both academia and industry, results could differ when employing other programming languages, LLMs with varying model sizes, or alternative prompting approaches. To mitigate these limitations, we employed a robust experimental design with a large-scale dataset comprising over 500k code samples, including human-written code sourced from approximately 17k real-world GitHub repositories, ensuring broad coverage and representativeness. Nonetheless, investigating additional configurations remains an important direction for future work to further validate and expand our conclusions.


\section{Conclusion}
\label{sec:conclusion}
This study offers a large-scale comparison of human-written and AI-generated code on multiple dimensions of software quality, including defects, vulnerabilities, and structural complexity. By applying standardized taxonomies, \ie ODC and CWE, we uncover fundamental differences in how code is produced by humans versus AI code assistants.
Human-written code tends to be more structurally complex but prone to maintainability and design issues, whereas AI-generated code is simpler but more repetitive and semantically shallow, with frequent unused constructs and hardcoded debugging. Critically, AI-generated code also exhibits a higher prevalence of high-severity security vulnerabilities, underscoring the need for stricter safeguards in AI-assisted development workflows.

By systematically exposing these differences and structuring them within a consistent evaluative framework, we believe that this study lays the groundwork for developing a standardized methodology that better reflects the quality and risk profiles of AI-generated code when compared to human-authored programs. 


\section*{Acknowledgment}
This work has been partially supported by the \textit{IDA—Information Disorder Awareness} Project funded by the European Union-Next Generation EU within the SERICS Program through the MUR National Recovery and Resilience Plan under Grant PE00000014, and the \textit{SERENA-IIoT} project funded by MUR (Ministero dell’Università e della Ricerca) and European Union (Next Generation EU) under the PRIN 2022 program (project code 2022CN4EBH).

\bibliographystyle{IEEEtran}
\bibliography{biblio.bib}

\end{document}